\documentclass[useAMS,usenatbib]{mn2e}
\usepackage{multirow}
\usepackage{epsfig}
\usepackage{subfigure}
\usepackage{url}

\makeatletter
\let\jnl@style=\rm
\def\ref@jnl#1{{\jnl@style#1}}

\def\aj{\ref@jnl{AJ}}                   
\def\araa{\ref@jnl{ARA\&A}}             
\def\apj{\ref@jnl{ApJ}}                 
\def\apjl{\ref@jnl{ApJ}}                
\def\apjs{\ref@jnl{ApJS}}               
\def\ao{\ref@jnl{Appl.~Opt.}}           
\def\apss{\ref@jnl{Ap\&SS}}             
\def\aap{\ref@jnl{A\&A}}                
\def\aapr{\ref@jnl{A\&A~Rev.}}          
\def\aaps{\ref@jnl{A\&AS}}              
\def\azh{\ref@jnl{AZh}}                 
\def\baas{\ref@jnl{BAAS}}               
\def\jrasc{\ref@jnl{JRASC}}             
\def\memras{\ref@jnl{MmRAS}}            
\def\mnras{\ref@jnl{MNRAS}}             
\def\pra{\ref@jnl{Phys.~Rev.~A}}        
\def\prb{\ref@jnl{Phys.~Rev.~B}}        
\def\prc{\ref@jnl{Phys.~Rev.~C}}        
\def\prd{\ref@jnl{Phys.~Rev.~D}}        
\def\pre{\ref@jnl{Phys.~Rev.~E}}        
\def\prl{\ref@jnl{Phys.~Rev.~Lett.}}    
\def\pasp{\ref@jnl{PASP}}               
\def\pasj{\ref@jnl{PASJ}}               
\def\qjras{\ref@jnl{QJRAS}}             
\def\skytel{\ref@jnl{S\&T}}             
\def\solphys{\ref@jnl{Sol.~Phys.}}      
\def\sovast{\ref@jnl{Soviet~Ast.}}      
\def\ssr{\ref@jnl{Space~Sci.~Rev.}}     
\def\zap{\ref@jnl{ZAp}}                 
\def\nat{\ref@jnl{Nature}}              
\def\iaucirc{\ref@jnl{IAU~Circ.}}       
\def\aplett{\ref@jnl{Astrophys.~Lett.}} 
\def\apspr{\ref@jnl{Astrophys.~Space~Phys.~Res.}}
\def\bain{\ref@jnl{Bull.~Astron.~Inst.~Netherlands}}
\def\fcp{\ref@jnl{Fund.~Cosmic~Phys.}}  
\def\gca{\ref@jnl{Geochim.~Cosmochim.~Acta}}   
\def\grl{\ref@jnl{Geophys.~Res.~Lett.}} 
\def\jcp{\ref@jnl{J.~Chem.~Phys.}}      
\def\jgr{\ref@jnl{J.~Geophys.~Res.}}    
\def\jqsrt{\ref@jnl{J.~Quant.~Spec.~Radiat.~Transf.}}
\def\memsai{\ref@jnl{Mem.~Soc.~Astron.~Italiana}}
\def\nphysa{\ref@jnl{Nucl.~Phys.~A}}   
\def\physrep{\ref@jnl{Phys.~Rep.}}   
\def\physscr{\ref@jnl{Phys.~Scr}}   
\def\planss{\ref@jnl{Planet.~Space~Sci.}}   
\def\procspie{\ref@jnl{Proc.~SPIE}}   

\makeatother

\title[A \textit{Chandra} view of the clumpy reflector at the heart of the Circinus galaxy]{A \textit{Chandra} view of the clumpy reflector at the heart of the Circinus galaxy}

\author[Andrea Marinucci, et al.]{A. Marinucci$^{1,2}$\thanks{E-mail: marinucci@fis.uniroma3.it (AM)}, G. Miniutti$^{2}$, S. Bianchi$^{1}$, G. Matt$^1$, G. Risaliti$^{3,4}$ \\
$^1$Dipartimento di Matematica e Fisica, Universit\`a degli Studi Roma Tre, via della Vasca Navale 84, 00146 Roma, Italy\\
$^2$ Centro de Astrobiolog\'ia (CSIC-INTA), Dep. de Astrof\'isica; LAEFF, Villanueva de la Ca$\tilde{n}$ada, Madrid, Spain\\
$^3$Harvard-Smithsonian Center for Astrophysics, 60 Garden St., Cambridge MA 02138, USA\\
$^4$INAF - Osservatorio Astrofisico di Arcetri, L.go E. Fermi 5, Firenze, Italy\\
}

\begin{document}
\maketitle
\label{firstpage}

\begin{abstract} 
We present a spectral and imaging analysis of the X-ray reflecting structure at the heart of the Circinus galaxy, investigating the innermost regions surrounding the central black hole. By studying an archival 200 ks Chandra ACIS-S observation, we are able to image the extended clumpy structure responsible for both cold reflection of the primary radiation and neutral iron K$\alpha$ line emission.  We measure an excess of the equivalent width of the iron K$\alpha$ line which follows an axisymmetric geometry around the nucleus on a hundred pc scale. Spectra extracted from different regions confirm a scenario in which the dominant mechanism is the reflection of the nuclear radiation from Compton-thick gas. Significant differences in the equivalent width of the iron K$\alpha$ emission line (up to a factor of 2) are found. It is argued that these differences are due to different scattering angles with respect to the line of sight rather than to different iron abundances. 
\end{abstract}
\section{ Introduction}
Seyfert 2 galaxies are a particular class of AGN that are fundamental for our understanding of the circumnuclear environment, due to their heavy obscuration.  If the absorbing material is thick to Compton-scattering (i.e. $N_{\rm H}\geq \sigma_{\rm T}^{-1} = 1.5 \times 10^{24}$cm$^{-2}$) the nuclear emission below 10 keV is completely obscured, allowing a clear view of components that are instead heavily diluted in unobscured sources, often down to invisibility.
The Circinus galaxy, one of the closest AGN \citep[D=$4.2^{+0.8}_{-0.8}$ Mpc:][]{free77} and brightest Seyfert 2 galaxies in the sky, has been largely observed by X-ray satellites in the last 20 years. It was observed by \textit{ROSAT} during the All Sky Survey for the first time in X-rays \citep{brink94} and later on by ASCA, showing a spectrum dominated by a pure Compton reflection component \citep{matt96}, with a prominent iron K$\alpha$ emission line and several other lines from lighter elements \citep{bmi01}. \textit{Beppo}SAX, some years later, confirmed the ASCA results below 10 keV and showed an absorbing column density of 4$\times10^{24}$cm$^{-2}$ \citep{matt99, gua99}. The properties of the nuclear emission studied with better angular resolution X-ray satellites such as \textit{Chandra}  \citep{Sambruna01a, Sambruna01b, mingo12} and XMM-\textit{Newton} \citep{mbm03,mass06} have been widely described. The extranuclear activity of Circinus has been very well studied too, including an [\textsc{OIII}] ionization cone \citep{marconi94}, two starburst rings at $\sim2$ arcsec and 10 arcsec from the nucleus \citep[][references therein]{wil00} and an overall complex extended radio structure \citep[][and references therein]{elmouttie98, curran08}. We refer the reader to \citet{maiolino01, ohsuga01} for a more extensive description of the complex nuclear structure.\\
In this work we present the analysis of an archival 200 ks {\em Chandra} observation, with the aim of investigating the innermost regions (from tens to hundreds of parsecs from the central nucleus) by constructing a map of the equivalent width (EW) of the Fe K$\alpha$ emission. We will show that such procedure is crucial to study the reprocessing of the nuclear radiation by the circumnuclear Compton-thick material, unveiling the dimensions and the geometrical structure of the absorber/reflector. 

\section{Observations and data reduction}\label{obsred}
The Circinus galaxy (z=0.00144) was observed by {\em Chandra} on 2010, December 17$^{\rm th}$ for an exposure time of 160 ks and a week later, on December 24$^{\rm th}$ for an exposure time of 40 ks, with the Advanced CCD Imaging Spectrometer \citep[ACIS:][]{acis}. Data were reduced with the Chandra Interactive Analysis of Observations \citep[CIAO:][]{ciao} 4.4 and the Chandra Calibration Data Base (CALDB) 4.4.6 database, adopting standard procedures. The imaging analysis was performed applying the Subpixel Event Repositioning and smoothing procedures widely discussed in the literature \citep[]{tsunemi01, li04, wang11b}. We therefore used a pixel size of 0.123 arcsec, instead of the native 0.495 arcsec. At the distance of the source 1 arcsec=19 pc. We generated event files for the two observations separately with the CIAO tool \textsc{chandra\_repro} and merged them into a single image using the tool \textsc{merge\_all}. After cleaning for background flaring we got a total exposure of 192 ks. Then we identified point sources in the field of view with the \textsc{wavdetect} tool and removed them, with the exception of the nucleus and a faint source detected by \textsc{wavdetect} in the 1.0-3.0 keV image (with the applied subpixeling factor of 4), 0.038 arcmin from the nucleus (RA=14:13:09.598; DEC=-65:20:21.85).\\
\begin{figure*}
 \includegraphics[width=\columnwidth]{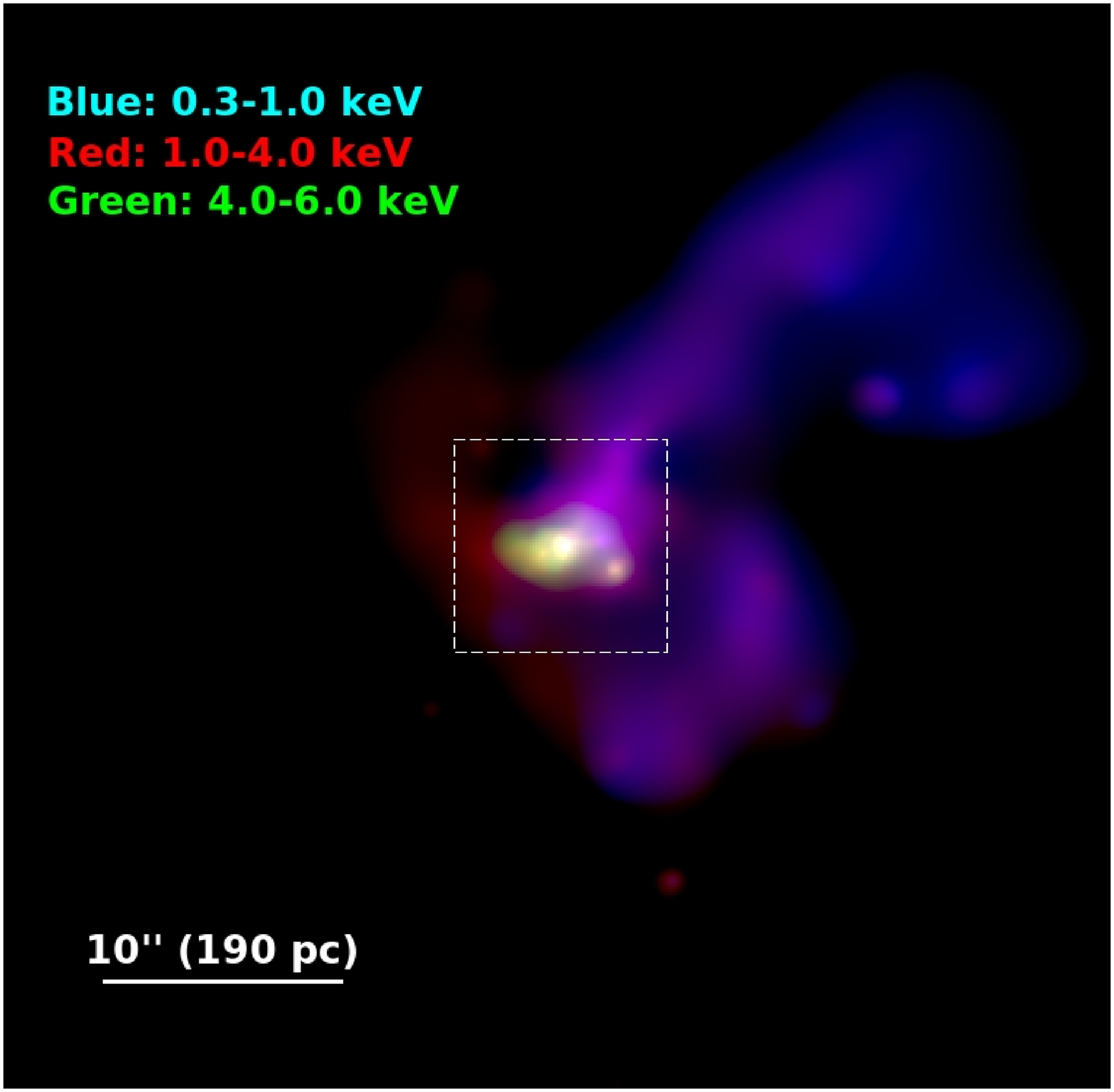}
    \includegraphics[width=1.015\columnwidth]{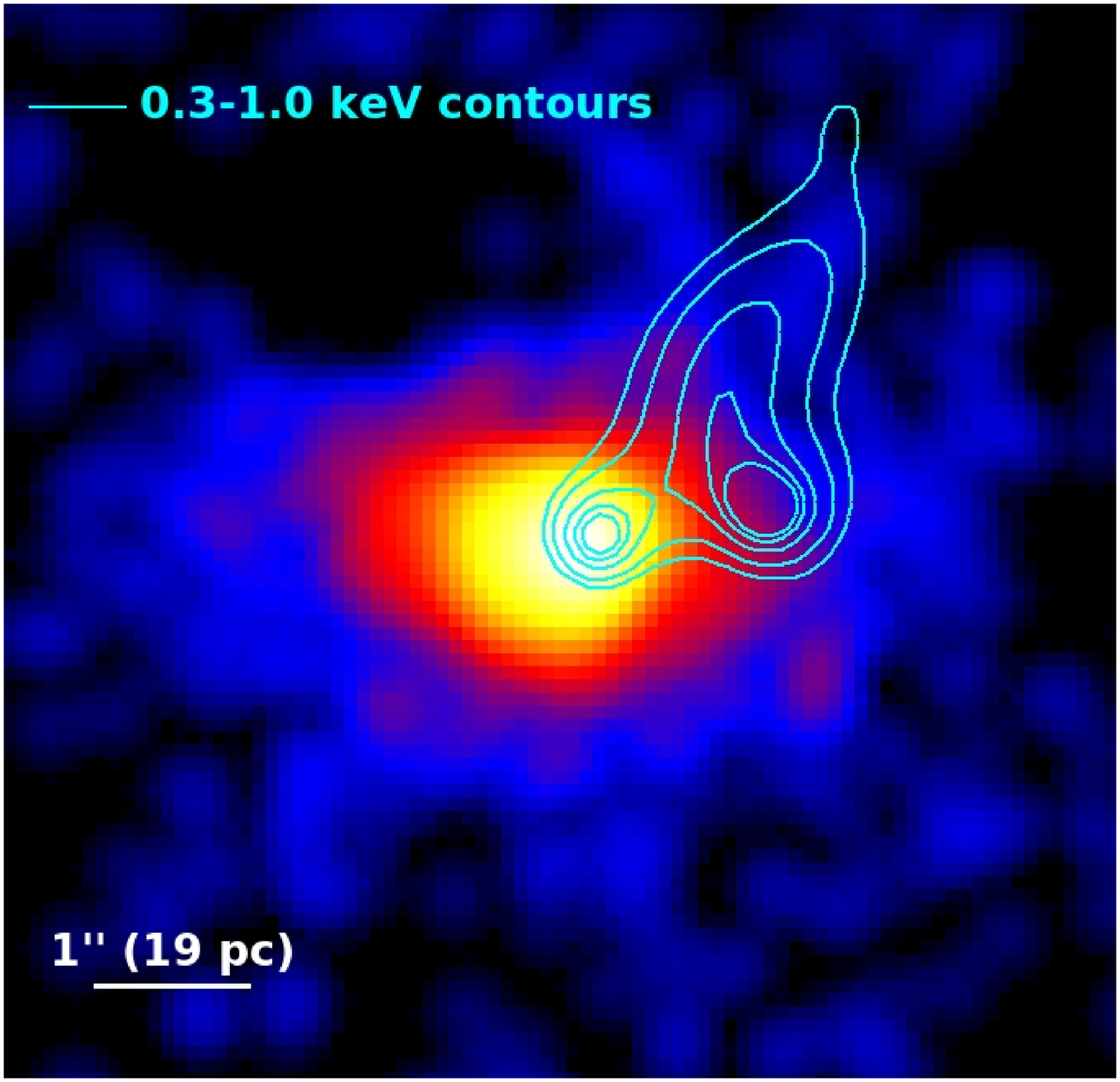}
  \caption{\textit{Left panel:} RGB image of the nuclear region in the Circinus galaxy. Several of the X-ray components extensively discussed in the past, such as the [\textsc{oiii}] ionization cone, the starburst ring and the compact nuclear reflected emission, can be clearly identified. \textit{Right panel:} Image of the emission between 6.3-6.5 keV with the over-imposed contours of the soft emitting cone.}
  \label{rgb}
\end{figure*}
Spectra 1, 2, 3, 4 were extracted from 1.5$\times$1.5 arcsec$^2$ regions, 5 and 6 from $3.5$ arcsec $\times$ 1 arcsec boxes and 7 from a 0.9$\times$0.9 arcsec$^2$ region. We used a $15$ arcsec radius circle for background extraction. Spectra were binned in order to over-sample the instrumental resolution by a factor of 3 and to have no less than 30~counts in each background-subtracted spectral channel. This allows the applicability of the $\chi^2$ statistics.\\
We ignored channels between 7.0 and 10.0 keV due to pileup, constrained to be within 10\%. We applied a self-consistent model of pileup  \citep[following the procedure described in][]{davis01} to the broadband 0.5-12 keV spectrum, extracted from the central 1 arcsec region, and then verified that our results below 7 keV are not significantly affected by pileup.
\begin{figure}
  \includegraphics[width=\columnwidth]{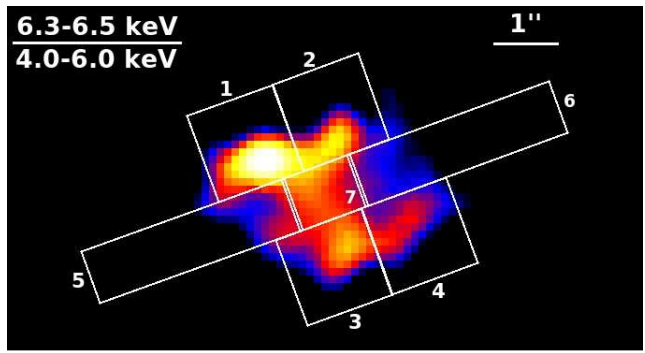}
  \caption{Zoomed image of the ratio between the 6.3-6.5 keV and the 4.0-6.0 keV emission. The excess of Fe K$\alpha$ EW is clumpy and axisymmetric with respect to the nucleus.}
  \label{line}
\end{figure}
\section{Data analysis}
\subsection{Imaging}
\begin{figure*}
  \includegraphics[width=\columnwidth]{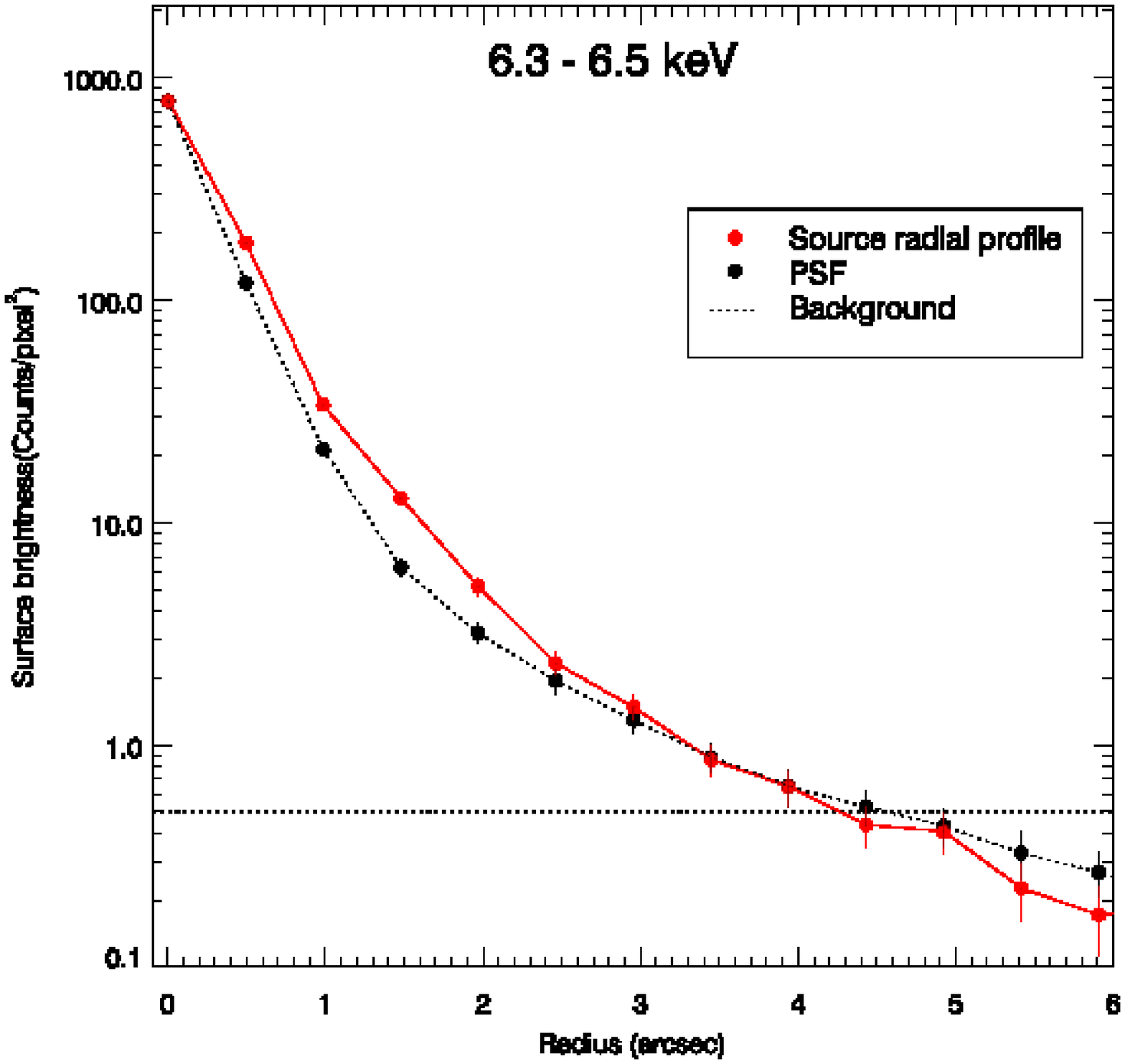}
  \includegraphics[width=\columnwidth]{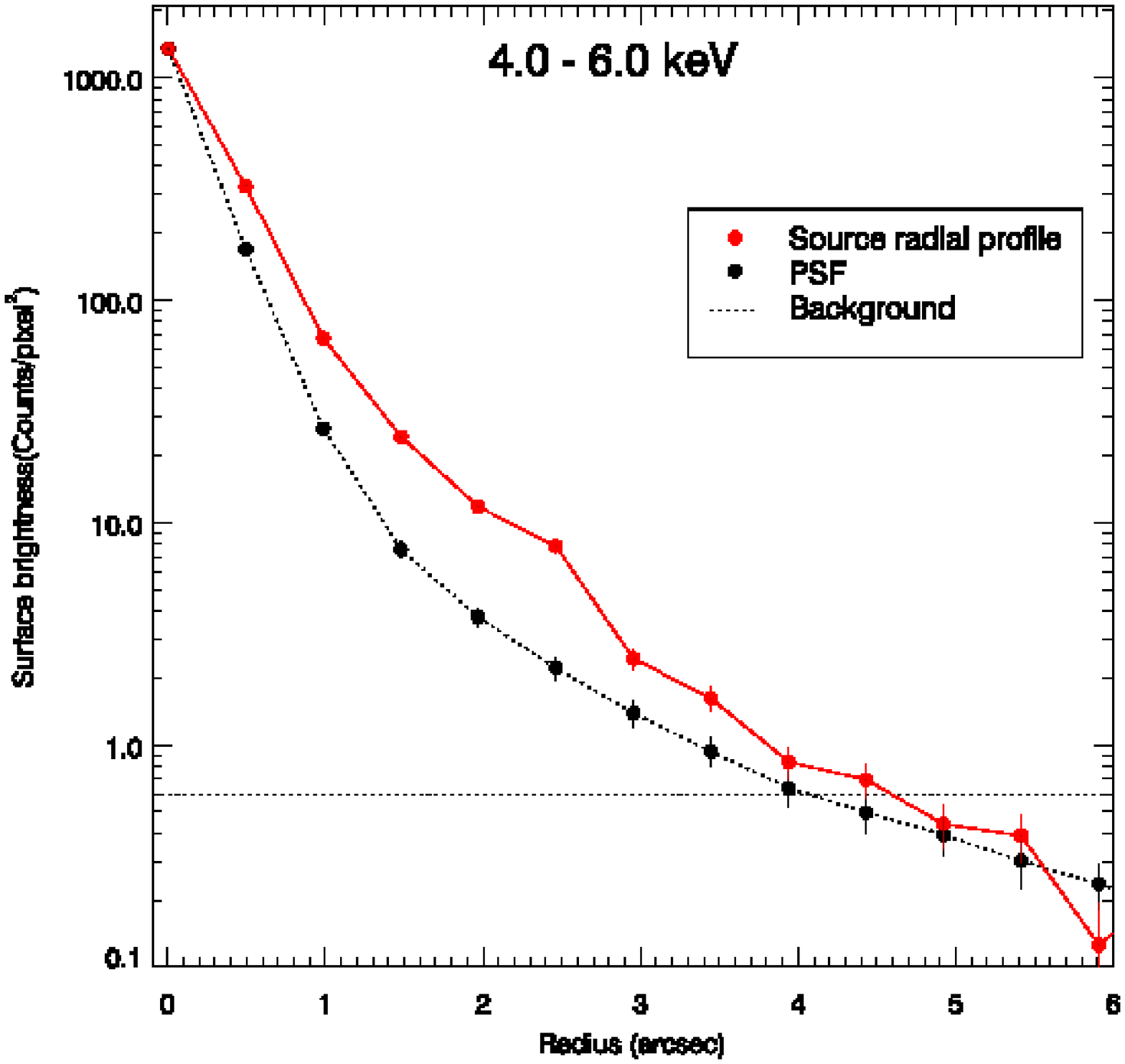}
  \caption{\textit{Left panel:} Radial profiles of the source and ACIS-S Point Spread Function in the 6.3-6.5 keV energy band. \textit{Right panel:}  Radial profiles of the source and ACIS-S Point Spread Function in 4.0-6.0 keV energy band. We chose 1 pixel radius, concentric annuli for the source and PSF surface brightness, and a 25 pixel radius annulus for the background one. The PSF radial profile was normalized to the one of the source.}
  \label{radprof}
\end{figure*}

In Fig. \ref{rgb} (left panel) the 50 arcsec $\times$ 50 arcsec central region is shown. The merged event file was filtered in energy between 0.3-1.0 keV (Blue), 1.0-4.0 keV (Red) and 4.0-6.0 keV (Green). Several components of the circumnuclear environment can be clearly identified in Fig. \ref{rgb} (left panel). While the cone of highly ionized gas emitting in the 0.3-1.0 keV band \citep{smithwilson01} perfectly resembles the optical [OIII]/(H$\alpha$+[NII]) image, the circumnuclear starburst ring, attributed to supernova remnants activity, matches the [SII]/H$\alpha$ image \citep[][Fig. 6 and 7, respectively]{marconi94}. A compact, intense hard X-ray emitting zone is apparent in the inner 200 parsec. It can be attributed to the reprocessing of the nuclear radiation by the obscuring Compton-thick material, on a hundred parsec scale from the central X-ray source.  In a radius of 3 arcsec, the unresolved nucleus contributes 70 per cent of the 0.5-8 keV counts, with the extended component contributing of the remaining counts \citep{Sambruna01a}. It is interesting to note that such spatial scale is consistent with the so-called ``obscuring wall'' region described in \citet{ohsuga01}, a radiatively-supported circmunuclear obscuring structure on scales $< 100$ pc. \\
To investigate this region in greater detail we therefore selected two energy intervals: one to characterize the neutral iron K$\alpha$ emission (6.3-6.5 keV, Fig. \ref{rgb}, right panel) and one for the reflected emission from cold material (4.0-6.0 keV). Such a conservative choice on the iron K$\alpha$ line energy interval was taken to avoid any contribution from further more ionized nuclear components, such as Fe \textsc{xxv} K$\alpha$ and Fe \textsc{xxvi} K$\alpha$. The soft emission, on the other hand, can be neglected above 4 keV, being the cold reflection the dominant spectral component. We then divided the 6.3-6.5 keV image by the 4.0-6.0 keV one and smoothed the image with a Gaussian filter with a kernel of 4 pixels: the result can be seen in Fig. \ref{line}. The Fe K$\alpha$ EW excess is axisymmetric with respect to the nucleus, co-aligned with the disk of the galaxy and almost perpendicular to the [\textsc{OIII}] ionization cone. \\ The structure is extended on a spatial scale that can be fully embedded in a 100$\times$100 parsecs region. Radial profiles are shown in Fig. \ref{radprof}, both in  the iron K$\alpha$ and in the cold reflection energy bands. The contribution of the source to the total surface brightness is in excess to the one expected from the ACIS-S Point Spread Function, up to $3$ arcsec ($\sim60$ pc). \\
\subsection{Spectral fitting}\label{spec}
\begin{figure}
  \includegraphics[ angle=-90, width=\columnwidth]{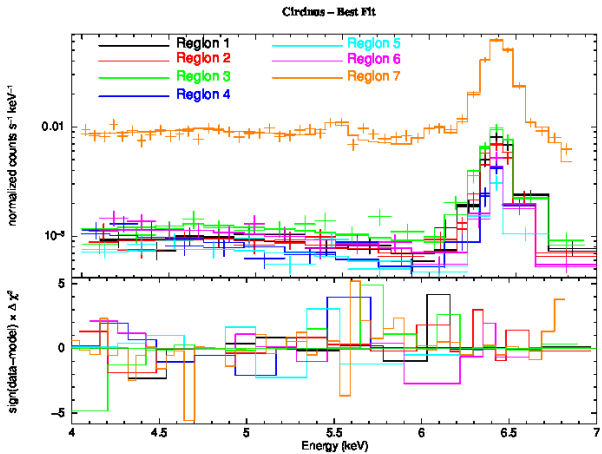}
  \caption{Best fits (top panel) and residuals (bottom panel) for the spectra extracted from the regions presented in Fig. \ref{line} and Table \ref{bestfit}. The main components of the model are the reflection from neutral material and the intense Iron K$\alpha$ emission line: no strong, combined residuals are present. See Sect. \ref{spec} for further details.}
  \label{fits}
\end{figure}

The 4-7~keV spectra of the 7 regions defined in Fig. \ref{line} are shown in Fig. \ref{fits}. We used \textsc{xspec 12.8.0} \citep{xspec} and errors correspond to the 90\% confidence level for one interesting parameter ($\Delta\chi^2=2.7$), if not otherwise stated.  The 7 spectra were fitted with a model consisting of a reflection continuum \citep[model PEXRAV,][]{mz95} and three emission lines for neutral Fe K$\alpha$, Fe \textsc{xxv} K$\alpha$ (actually an unresolved triplet, we fixed the centroid of the line at 6.65 keV) and for the Compton Shoulder (CS) redwards of the line core \citep[with energy fixed at 6.3 keV and $\sigma=40$ eV, ][]{matt02}.  An additional power law was added to the model, when needed, to be associated to the diffuse soft emission.A value of $\Gamma=1.56$ was used for the reflection component, as measured from the BeppoSAX and Suzaku high energy spectra \citep{matt99, yangwi09}, and the cosine of the inclination angle is fixed to 0.45. A Galactic absorption of $5.6 \times 10^{21}$cm$^{-2}$ has been taken into account \citep{dl90}.  We left the normalizations of the emission lines, of the reflection component, the energy centroid and width of the Fe K$\alpha$ line free to vary between the 7 spectra. \\
The model successfully reproduces all the individual spectra, with EW typical of reflection-dominated sources. The overall fit is good and no strong residuals are present. The additional power law ($\Gamma=3.0\pm0.2$) for the diffuse emission is required only in the fit of the spectrum extracted from region 6 ($\Delta \chi^2=5$), with a contribution smaller than 10 per cent to the total flux in the 4-7 keV band. In region 4 a power law ($\Gamma=1.6\pm0.2$) is required ($\Delta \chi^2=8$) and can be associated to the faint source mentioned in Sect. \ref{obsred}, its flux contributing $\sim$15 per cent of the total. In region 7 two additional emission lines are also required by the fit, at $5.45\pm0.05$ keV and $6.03\pm0.04$ keV ($\Delta \chi^2=8$) with fluxes of $0.2\pm0.1$ ad $0.3\pm0.1$ $10^{-5}$ ph cm$^{-2}$ s$^{-1}$, respectively. These two emission lines had already been detected in the grating spectra presented in \citet{Sambruna01b}. The best fit leads to a $\chi ^2/$d.o.f.=120/102 and no combined additional residuals are found (see Fig. \ref{fits}).\\
When we measure the EW of the narrow core of the neutral iron K$\alpha$ line we find a clear variation between different zones, as inferred from the axisymmetric structure shown in Fig. \ref{line}. The maximum variation, at a 2.4 $\sigma$ confidence level, is found when we compare region 1 (EW=$2.2^{+0.5}_{-0.5}$ keV) with region 5 (EW=$0.9^{+0.3}_{-0.3}$ keV).
\begin{table*}
\begin{center}
\begin{tabular}{cccccccc}
& & & & & & & \\
\hline
\hline
{\bfseries Region} &{\bfseries Counts}   &{\bfseries F$_{\rm Fe\ XXV K\alpha}$ } & {\bfseries CS} & {\bfseries E$_{\rm C}$  } &{\bfseries F$_{\rm Fe\ K\alpha}$ }  & {\bfseries EW} &{\bfseries F$_{\rm 4-10\ keV }$ }     \\
\hline
 {\bfseries 1}& 775 &$0.2^{+0.1}_{-0.1}$& $<0.25$&$6.41^{+0.01}_{-0.02}$& $0.77^{+0.18}_{-0.10}$& $2.2^{+0.5}_{-0.5}$& $2.0^{+0.1}_{-0.1}$\\
{\bfseries 2} &714 &  $0.10^{+0.05}_{-0.05}$&$<0.25$ & $6.42^{+0.01}_{-0.02}$& $0.65^{+0.15}_{-0.10}$& $2.0^{+0.5}_{-0.5}$& $1.7^{+0.1}_{-0.1}$\\
{\bfseries 3}& 974& $0.14^{+0.08}_{-0.10}$ & $<0.4$ & $6.42^{+0.01}_{-0.01}$&$0.83^{+0.17}_{-0.12}$ & $1.9^{+0.5}_{-0.4}$ & $2.3^{+0.1}_{-0.1}$\\
{\bfseries 4} & 516&$<0.4$ & $<0.15$& $6.42^{+0.01}_{-0.03}$& $0.4^{+0.1}_{-0.1}$& $1.6^{+0.6}_{-0.5}$& $1.1^{+0.3}_{-0.3}$\\
{\bfseries 5}& 432& $<0.16$& $<0.07$& $6.38^{+0.02}_{-0.02}$& $0.26^{+0.07}_{-0.07}$& $0.9^{+0.3}_{-0.3}$& $1.0^{+0.1}_{-0.1}$\\
{\bfseries 6}& 614&$<0.12$ &$<0.08$ &$6.41^{+0.01}_{-0.01}$ & $0.4^{+0.1}_{-0.1}$& $1.0^{+0.3}_{-0.2}$&$1.3^{+0.1}_{-0.1}$ \\
{\bfseries 7}& 6839& $0.6^{+0.1}_{-0.1}$& $1.0^{+0.2}_{-0.2}$&$6.417^{+0.005}_{-0.005}$ & $6.1^{+0.4}_{-0.4}$& $1.8^{+0.2}_{-0.2}$&$16.2^{+0.4}_{-0.4}$ \\
\hline
\hline
\end{tabular}\\
\caption{\label{bestfit} Best-fitting parameters for the spectra extracted from regions in Fig. \ref{line}. Energies and EWs are in keV, line fluxes in $10^{-5}$ ph cm$^{-2}$ s$^{-1}$, observed fluxes in  $10^{-13}$ erg cm$^{-2}$ s$^{-1}$. See text for details.}
\label{bestfit}
\end{center}
\end{table*}

\section{Discussion}
In the last few years much evidence for the presence of extended iron K$\alpha$ emission has been found in nearby Compton-thick AGN, such as NGC 1068 \citep{yws01}, Mrk 3 \citep{guainazzi12} and NGC 4945 \citep{mari12b}, indicating that reflecting material may extend on scales ranging from tens to thousands of parsecs. We show with our analysis that high resolution X-ray imaging is a very powerful tool to trace this component and it is fundamental to disentangle the extended reflecting structures from the nuclear unresolved emission. The map of the EW of the line with respect to the cold reflection continuum allows us to draw a general geometry for the circumnuclear environment, on spatial scales of tens of parsecs. The gas we picture in Fig. \ref{rgb} is aligned with the one shown in a recent catalog of mid-infrared observations of Circinus \citep[][references to previous works therein]{bumeitri13} where smaller scales are studied ($\sim$0.1-10 pc), suggesting that the material producing the cold reflection and the iron K$\alpha$ line, in principle, could be the outer part of the dusty region responsible for the infrared emission. \\
The maximum difference in EW is found when regions 5 and 1 are compared (a factor $\sim2$). A scenario where such a difference is due to variations of metallicity between the two zones is unlikely. The dependence of the Fe K$\alpha$ features from the chemical abundances is discussed in \citet{mfr97} and a difference of 2-5 in iron abundance is needed to achieve a difference of a factor 2 in the EW of the line. 
Enrichment due to iron is mainly attributed to SNIa on times scales of $\sim$1 Gyr (SNIa late) or hundreds of Myrs (SNIa early) \citep[][and references therein]{loewe06}. Such an iron enrichment on these time scales cannot remain constrained in a region of tens of parsecs. A difference of 3-5 Z$_{\odot}$ between regions 1 and 5 can therefore be considered very unlikely.\\
We interpret the differences in the EW of the iron K$\alpha$ line (Table \ref{bestfit}) and the particular, axisymmetric shape of its excess with respect of the reflected nuclear continuum (Fig. \ref{line}) as purely due to geometrical effects. The bright Fe line EW spots (e.g. regions 1 and 3 in Fig \ref{line}) are regions where Fe emission is enhanced with respect to the underlying reflection continuum.  Indeed, the Fe K$\alpha$ EW has a strong dependence from the column density of the illuminated material, as presented in a large number of works using a toroidal \citep{yaqoob10} or a slab \citep{matt02} geometry, and from the angle $\theta_i$ between the polar direction and the line of sight, as studied and discussed in \citet{mpp91} and \citet{gf91}.  The axisymmetric nature of the Fe line EW image seen in Fig. \ref{line} suggests that the two brightest EW spots (regions 1 and 3) define the equatorial plane of the obscuring medium and accretion flow system, aligned with the disk of the galaxy and with the structure observed in mid-infrared \citep{bumeitri13}. Reassuringly, the normal to the direction joining regions 1 and 3 is aligned with the soft X-rays (and [OIII] cones), which most likely define the polar direction.\\
We point out the fact that the soft emitting cone is observed (Fig. \ref{rgb}) in regions where also iron K$\alpha$ is emitted: it is a further hint of the presence of a clumpy structure \citep{nenkova08} due to soft emission leaking through the reflecting material, as already found in similar imaging analysis with Chandra on NGC 4151 \citep{wang11b} and Mrk 573 \citep{paggi12}.

\section{Conclusions}
We presented a spectral and imaging analysis of the X-ray reflecting structure at the center of the Circinus galaxy, taking advantage of a 200 ks Chandra ACIS-S observation. An axisymmetric, clumpy structure is responsible for both cold reflection of the primary radiation and neutral iron K$\alpha$ line emission, on spatial scales of tens of parsecs. Spectra extracted from different regions can be interpreted with cold reflection from Compton-thick gas and significant variations in the equivalent width of the iron K$\alpha$ emission line.ÊWe discard a scenario where the observed differences can be explained in terms of a metallicity variation, we conclude that they are likely explained in terms of scattering angles with respect to the line of sight. 

\section*{Acknowledgements}
We thank the anonymous referee for comments and suggestions. We thank Roberto Maiolino for discussions. AM acknowledges Fondazione Angelo Della Riccia for financial support. We would  also like to thank Kazushi Iwasawa and Francesco Massaro for suggestions. 

\bibliographystyle{mn2e}
\bibliography{sbs}

\end{document}